\documentclass{aa}  
\usepackage{graphicx}
\usepackage{color}
\usepackage{txfonts}
\usepackage{natbib}

\begin{document}
   \title{A novel L-band imaging search for giant planets \break in the Tucana and $\beta$ Pictoris moving groups
   \thanks{Based on observations collected at the European Southern Observatory, Chile, through the proposals 073.C-0834(A) and 074.C-0323(A).}}


   \author{M. Kasper\inst{1}
          \and D. Apai\inst{2,3}
          \and M. Janson\inst{4}
          \and W. Brandner\inst{4,5}
          }

   \offprints{M. Kasper, \email{mkasper@eso.org}}

   \institute{European Southern Observatory (ESO), Karl-Schwarzschild-Str. 2, D-85748 Garching
         \and Steward Observatory, The University of Arizona, 933 N. Cherry Avenue, Tucson, AZ-85721
         \and LAPLACE Team, NASA Astrobiology Institute
         \and Max-Planck-Institut f\"ur Astronomie, K\"onigstuhl 17, D-69117 Heidelberg
         \and UCLA, Division of Astronomy, Los Angeles, CA 90095, USA
             }

   \date{Received xx, 2006; accepted xx, xx}

 
  \abstract
	{Direct imaging using various techniques for suppressing the stellar halo nowadays achieves the contrast levels required to detect and characterize light of substellar companions at orbital distances larger than a few astronomical units from their host stars. The method nicely complements the radial velocity surveys that provide evidence that giant extrasolar planets in close-in orbits are relatively common.}  
   {The paper presents results from a small survey of 22 young, nearby stars that was designed to detect substellar companions and ultimately giant extrasolar planets down to Jupiter masses. The targets are members of the Tucana and $\beta$~Pictoris moving groups apart from the somewhat older star HIP~71395 that has a radial velocity trend suggesting a massive planet in large orbit.}
   {The survey was carried out in the L-band using adaptive optics assisted imaging with NAOS-CONICA (NACO) at the VLT. The chosen observation wavelength is well suited to search for close companions around young stars and delivers unprecedented detection limits. The presented technique reaches some of the best sensitivities as of today and is currently the most sensitive method for the contrast limited detection of substellar companions that are cooler than about 1000\,K.}
   {The companion to 51~Eri, GJ~3305, was found to be a very close binary on an eccentric orbit. No substellar companions were found around the target stars, although the method permitted to detect companions down to a few Jupiter masses at orbital distances of typically 5 astronomical units. A planet with a mass $\ge$1 M$_{\rm Jup}$ at distances $\ge$5\,AU around AU~Mic can be excluded at the time of our observations. The absence of detected planets sets constraints on the frequency distribution and maximum orbital distance of giant exoplanets. For example, a radial distribution power law index of 0.2 in combination with a maximum orbital radius exceeding 30 AU can be rejected at a 90\% confidence level.}
   

   \maketitle
%

\section{Introduction}

A major scientific breakthrough of the last years was the discovery of extra-solar planets by radial velocity searches \citep{mayor95} which add up to close to 200 such systems known to date (see www.exoplanets.org). \citet{butler06} discuss that even modest extrapolation of the currently available data suggests that a significant fraction of around 12\% of the nearby stars harbour giant planetary companions within 30\,AU, and that a large population of Jupiter mass planets beyond 3\,AU is probable. Further, the frequency of planets quickly drops with increasing planetary mass underlining the importance of a low mass detection limit for a dedicated imaging survey in order to improve the chances of success. However, the discovery space of radial velocity searches does not yet cover planets at distances larger than about 5\,AU, so few observational results on these objects exist to date. The recent detections of substellar \citep[e.g.][]{mccaughrean04,neuhaeuser05,biller06} or even planetary mass \citep{chauvin05} companions around young stars by direct imaging with Adaptive Optics (AO) let this technique appear most promising to close this gap. 

Previous surveys looking for substellar companions were mostly carried out in the near infrared either by broadband imaging \citep{chauvin03, neuhaeuser03, masciadri05, lowrance05} or using spectral differential techniques in narrow filters inside and outside methane absorption features to efficiently subtract speckle noise \citep{close05, biller07}. The thermal infrared instead has rarely been used so far. This wavelength range offers considerable advantages compared to shorter wavelengths, the most important one being the improved contrast of planetary mass companions with respect to their host stars. The models of \citet{baraffe03} predict H--L colors of around 2.5\,mag for 1000\,K hot companion (5 Jupiter mass object at the age of 30\,Myr or 30 M$_\mathrm{J}$ of 1\,Gyr) and H--L$\approx$4.5\,mag for a 350\,K warm companion (1M$_\mathrm{J}$ of 30\,Myr or 5 M$_\mathrm{J}$ of 1\,Gyr). Obviously the L-band offers tremendous gains for observing lower mass and older objects. As an example, the planetary mass companion to the brown dwarf 2M1207 is as red as H--L$\approx$3.81\,mag \citep{chauvin05}. An additional advantage of L-band over shorter wavelength observations is the better and more stable image quality delivered by AO reducing speckle noise and facilitating point spread function (PSF) subtraction. All these advantages more than compensate for the increased sky background in the thermal infrared. This technique has also been demonstrated on the Multiple Mirror Telescope with the Clio instrument \citep{hinz06,heinze06}.

Here we report the first L--band southern high--contrast imaging survey that probes giant planets at larger radii around young stars, a population, that is otherwise unconstrained by indirect methods such as the radial velocity, planetary transits and gravitational lensing. Such a giant planet population may be produced either through in--situ formation in massive disks, outward migration \citep{veras04}, or planet--planet scattering \citep[e.g.][]{ford05}. The survey obtained some of the most sensitive images of the vicinity of these young stars. We discuss the sample, the observational technique and the detection limits. Finally, we use a Monte Carlo simulation to constrain the properties of the radial distribution of giant planets.


\section{Targets and Observations}

The target sample has been compiled from members of the Tucana association \citep{zuckerman01b} and the $\beta$ Pic moving group \citep{zuckerman01a}. The ages of the member stars were estimated to be between 10 and 30\,Myr \citep[e.g.][]{kastner03, stelzer01} providing an ideal test bed for detecting extra-solar planets and study their formation. In selecting the targets, nearby and later type stars were prioritized in order to maximize contrast of a putative companion at a given angular separation. The final sample consists of 22 nearby (10-46\,pc) G, K, and M stars of which 10 are part of binary systems. Only HIP\,71395 (age between 0.5 and 1\,Gyr) does not belong to a young association and was added to the target list because of its proximity, comparatively young age and radial velocity trend suggesting a giant planet at larger orbital separation \citep{butler03}.
 
Table~\ref{tab:ts} summarizes the target list and provides information on the relevant parameters in the context of our survey. The L-band magnitudes have been estimated from reported V-band luminosities according to spectral types and their typical colors. This strategy was tested on some not saturated stars and was found to be accurate within about $\pm0.2$ magnitudes, a level of accuracy that was expected due to errors in V, spectral type, color, and possible unknown IR excess.

\begin{table*}
  \begin{center}
  \caption{Target stars from the Tucana and $\beta${} Pic (BPMG) moving groups.}
  \label{tab:ts}
  \begin{tabular}{lcccccl}\hline
       						& Spectral & Distance & V & L  & Age & \\
       Target 		&  type & (pc) & (mag) & (mag) & (Myr) & Remarks\\\hline
       HIP 71395  & K0 & 17	& 7.6	& 5.6	& 500-1000 & lin. radial velocity trend, observed in P73\\
       HIP 116748	& G6IV & 46	& 8.5	& 6.8	& 10-30	& Tucana, 5.3$^{\prime\prime}$ binary, P73\\
       HIP 102141	& M4.5 & 10	& 11.3 & 5.3 & 12	& BPMG, AT Mic, GJ799, 2.8$^{\prime\prime}$ binary, P73\\
       HIP 102409	& M0	& 10	& 8.9	& 5.1	& 12	& BPMG,	AU Mic GJ803, debris disk, P73\\
       HIP 107649	& G0V	& 16	& 5.6	& 4.2	& 10-30	& Tucana, P73\\
       HIP 105441a	& K2V	& 31	& 9	& 6.8	& 10-30	& Tucana, P73\\
       HIP 105441b & $>$K2	& 31	& 10.5	& 8	& 10-30	& Tucana, P73\\
       HIP 1993	& K7V	& 37	& 11.5	& 8.2	& 10-30	& Tucana, P73\\
       HIP 3556	& M1.5	& 39	& 11.9	& 7.9	& 10-30	& Tucana, P73\\
       HIP 2729	& K5V	& 46	& 9.6	& 6.7	& 10-30 & Tucana, P73\\
       HIP 4448	& K4V	& 44	& 9	& 6.3	& 10-30 & Tucana, P73\\
       HIP 1113	& G6V	& 44	& 8.8	& 7.2	& 10-30 & Tucana, P73\\
       HIP 1481	& G0V	& 41	& 7.5	& 6.1	& 10-30 & Tucana, P73\\
       PPM 366328	& K0	& $\approx$60	& 9.7	& 7.52 & 10-30	& Tucana, P74\\
       HIP 1910a	& M1	& 46	& 11.5	& 7.44	& 10-30	& Tucana, 0.65$^{\prime\prime}$binary, P74\\
       HIP 1910b	& $>$M1	& 46	& ?	& 8.84	& 10-30	& Tucana, P74\\
       GJ 3305a	& M1	& 30	& 10.6	& 6.5	& 12 & BPMG, 0.093$^{\prime\prime}$ binary, P74\\
       GJ 3305b	& $>$M1	& 30	& ?	& 7.38	& 12 & BPMG, P74\\
       HIP 23309	& M1	& 26	& 10	& 6.16	& 12 & BPMG, P74\\
       HIP 25486	& F6	& 27	& 6.3	& 5.27	& 12 & BPMG, P74\\
       HIP 29964	& K7	& 38	& 9.8	& 6.77	& 12 & BPMG, AO Men, P74\\\hline
  \end{tabular}
  \end{center}
\end{table*}

All targets have been observed in the L$^{\prime}$ filter centered at 3.8\,$\mu$m with the L27 camera of CONICA providing a pixel scale of 27 mas. For part of the data (VLT observing period 73, P73, April to September 2004), NAOS was counter-chopping, i.e. the field selector of the wavefront sensor followed the chopping motion of the telescope's secondary mirror. Counter-chopping provides AO corrected images at both chopping positions which effectively doubles the observing efficiency. Chopping frequencies were in the range 0.1-0.2 Hz. The P74 (October 2004 to March 2005) data were taken a using a standard dithering procedure and a 33 degrees angular rotation of NACO halfway through the observations. Ideally this operation, called angular differential imaging \citep[ADI,][]{marois06} or roll subtraction when used at the HST \citep{lowrance05}, would rotate the field and potential companions around the host star while leaving the residual instrumental speckle pattern unaffected. At an alt-az telescope as the VLT, however, this operation also rotates residual telescope aberrations which somewhat reduces its efficiency when compared to application at the HST. 

All targets were observed in pairs that were chosen to match their celestial positions and L-band luminosities most closely. In this way, the objects could be used to calibrate or subtract each others PSF without the expense of additional observing time for calibrator stars. Total on-source exposure times for all targets are in the range 15 to 20 minutes. 

\begin{figure}
 \centering
 \includegraphics[width=\columnwidth]{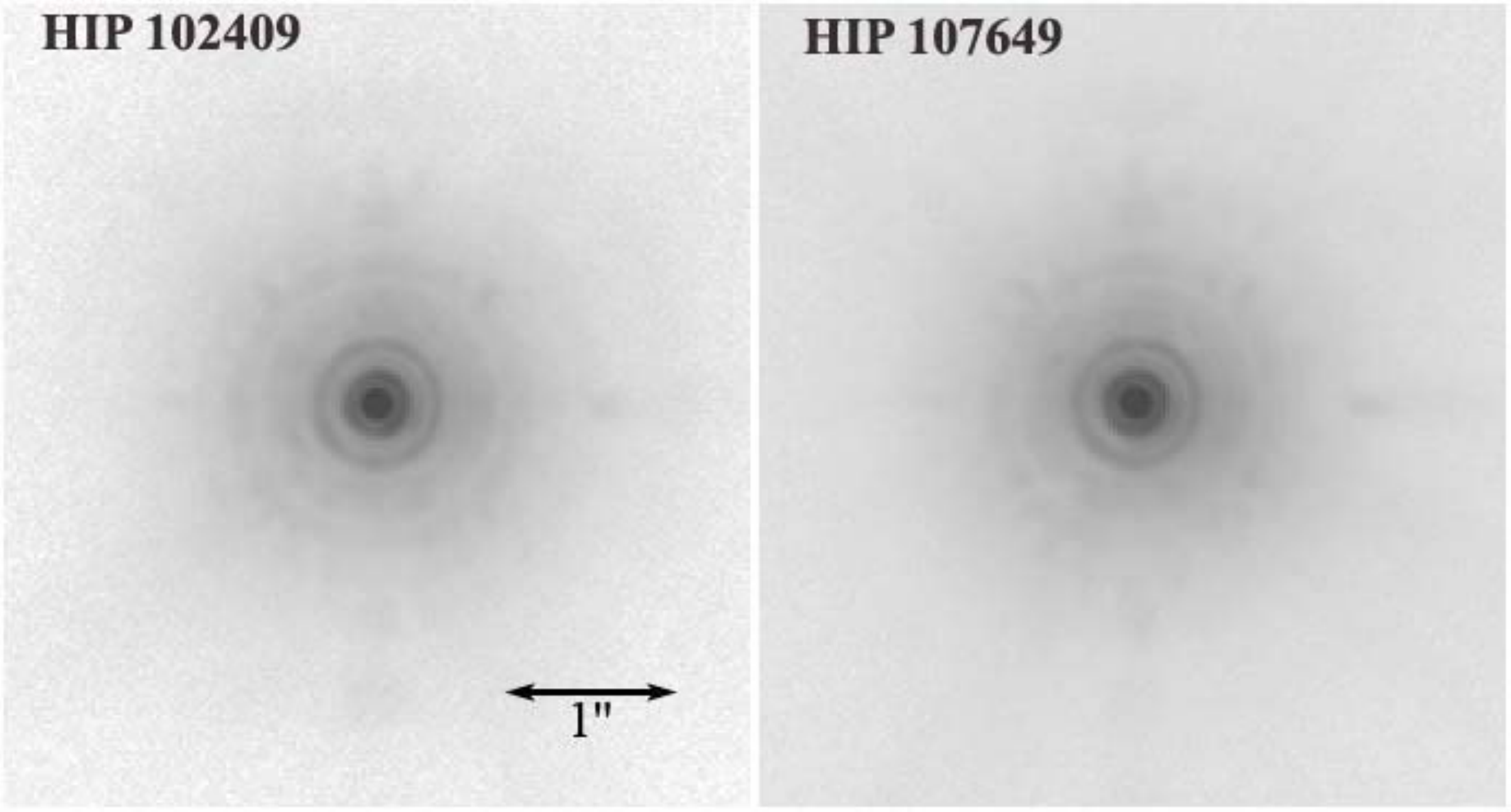}
 \caption{NACO L-band images of HIP 102409 and HIP 107649.}
 \label{fig:starfull}
\end{figure}
%

Figure~\ref{fig:starfull} shows HIP 102409 and HIP 107649 observed for 15 minutes on source each. The observations were separated by about 30 minutes from each other. The image quality is remarkably good with an L-band Strehl ratio of about 60-70\% which is typical for our data set. In addition, the stable image quality delivered by NACO at longer wavelengths greatly facilitates efficient PSF calibration and subtraction.

\section{Analysis}

\subsection{Data reduction}
\begin{figure}
 \centering
 \includegraphics[width=\columnwidth]{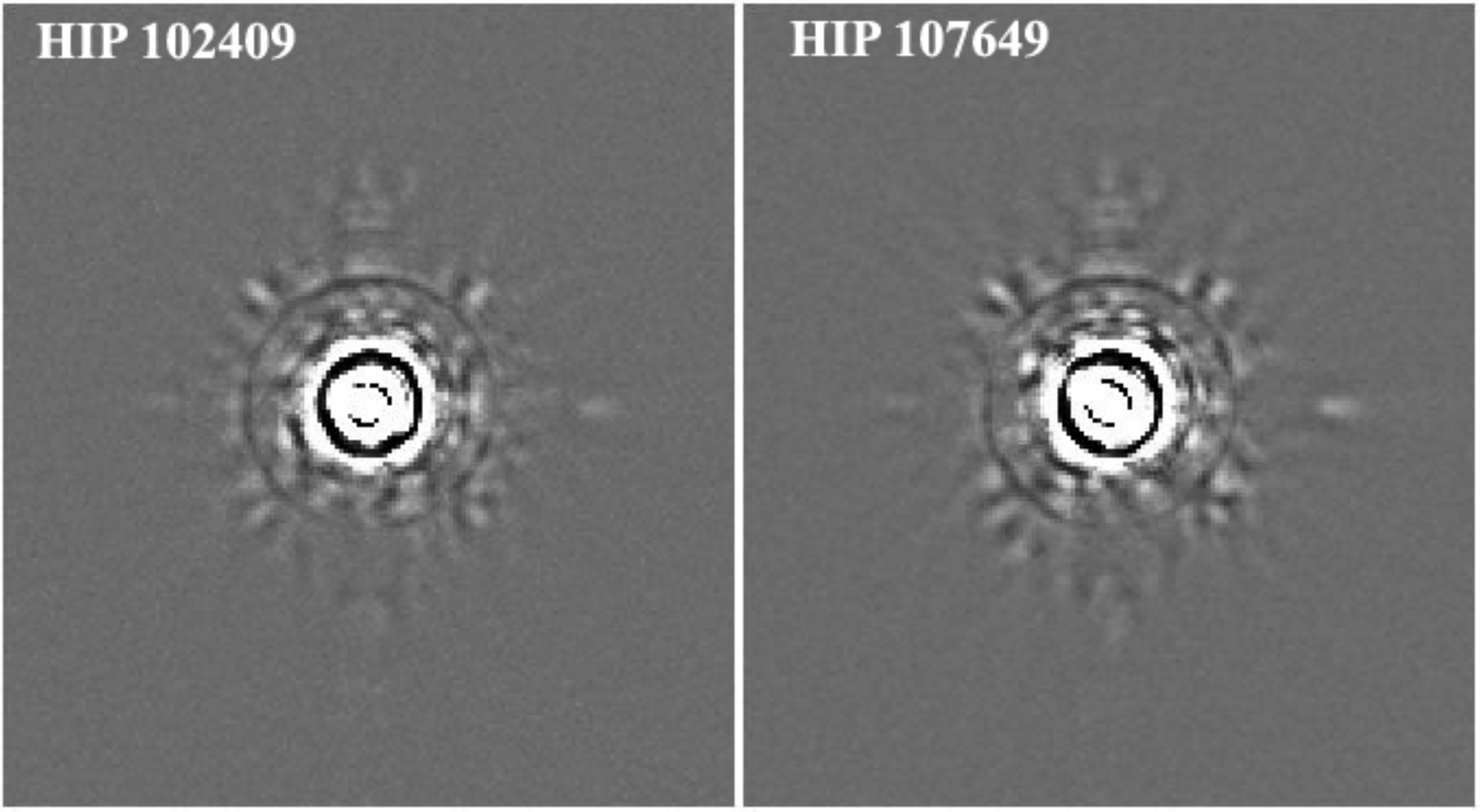}
 \caption{High-pass filtered images of HIP 102409 and HIP 107649.}
 \label{fig:starhf}
\end{figure}

After standard image processing (flat fielding, bad pixel cleaning, sky subtraction), the image pairs were superimposed with subpixel accuracy, flux normalized to unit peak intensity and high-pass filtered. The high-pass filtering was done by subtracting from the image its median filtered ($\approx$3 FWHM box-width) version. This procedure efficiently eliminates the large--scale structures while leaving point sources (such as substellar companions) relatively unaffected. The high-pass filtering reduces the PSF peak intensity by only about 15\%. Figure~\ref{fig:starhf} shows a high-pass filtered version of figure~\ref{fig:starfull}. The similarity of the two stars' residual speckle patterns is obvious. Finally the images have been median filtered with a small three pixels wide box in order to reduce pixel-to-pixel variations. 

The sky limiting magnitude (further away from the star and not affected by the contrast limit) imposed by stellar residuals has been estimated from the images after standard image processing. On average NACO reached a background limited point source sensitivity of about L = 16 to 16.5 mag at 5 sigma in 15 minutes on source integration time using counter chopping. The limiting magnitude was about 0.5 magnitudes brighter in dithering operation. However, the advantage in limiting magnitude when using counter chopping came at the expense of a somewhat less stable PSF due to occasional synchronization problems between AO and the telescope secondary mirror.

\subsection{Speckle reduction}
After high-pass filtering, residual speckle reduction for the P73 dataset was achieved by PSF subtraction. The P74 data set was observed with adaptor-rotator offsets of 33 degrees half way through the observation. Hence, the ADI technique was tested as an alternative to the PSF subtraction. ADI and PSF subtraction lead to very similar results in the final contrast suggesting that both equally well calibrate the quasi-static speckles resulting from imperfections of instrument optics. However, ADI offers the additional advantage that a planet would appear twice in the final image, a positive and a negative image would appear separated by the 33 degrees adaptor-rotator offset increasing the confidence level at which a potential planet would be considered as real. 

While the 26\arcsec{} wide binary HIP~105441 is reduced following the standard procedure, the medium separation binaries HIP~116748 (5.8\arcsec) and GJ~799 (2.8\arcsec) offer the unique possibility of simultaneous PSF subtraction. Since the anisoplanatic angle of the atmosphere is normally larger than half an arcminute in L-band and the light of both stars traverse almost the same instrument optics, the two components have nearly identical PSFs and efficiently calibrate each others speckle pattern in an SDI like fashion. A little problematic were the speckle calibration of the close binaries HIP~1910 (0.65\arcsec) and GJ~3305 (0.093\arcsec) whose PSFs overlap and therefore cannot be subtracted from each other. For these binaries, artificial PSF calibrators were created from the PSF reference stars by adding a shifted and properly flux scaled version of itself at the companion's position.

\begin{figure}
 \center{\includegraphics[width=2in,angle=0]{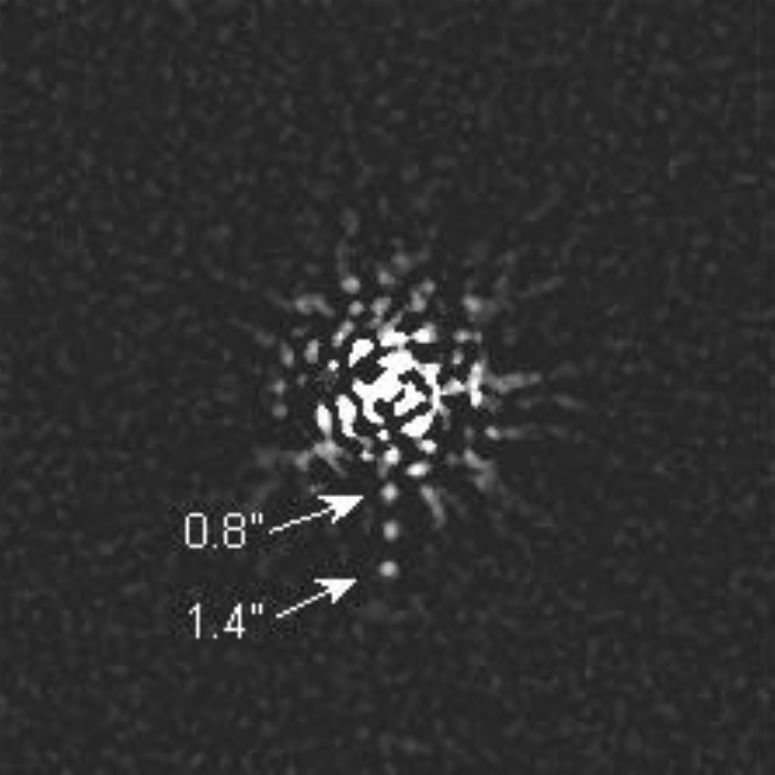}}
  \caption{Difference of HIP 102409 and HIP 107649 with artificial test planets 10 mag fainter than the stars.}\label{fig:diffima}
\end{figure}

Figure~\ref{fig:diffima} shows the difference of the two images shown in figure~\ref{fig:starhf} with fake planets 10 magnitudes fainter than the stars that have been added before the high-pass filtering at distances of 0.8\arcsec, 1.1\arcsec and 1.4\arcsec{}. All three planets are detected at the 5$\sigma$ level demonstrating the excellent sensitivity of the method. Since the planets are added before any filtering processes, this mandatory test would reveal any problem during data reduction and provides confidence in the predicted contrast levels.

\subsection{Detection limits}
The detection limit as a function of angular separation has been derived from the PSF or ADI subtracted images. At a given radial distance from the central star, the standard deviation of all pixel intensities in a four pixels (108 mas) wide annulus centered on the star provides the noise estimate. Since the mean flux of the subtracted image is near zero, this method accurately provides the variation of speckle intensity. Our method differs from the one used by \citet{biller07}, \citet{masciadri05} and \citet{neuhaeuser03} who calculated the standard deviation over a PSF size area for noise estimation. The latter method does not consider the mean intensity over this area as noise and therefore leads to contrast level estimates that are more optimistic than the ones delivered by our method. We tested both methods on our data and consistently found a difference of about 0.5 magnitudes in predicted contrast between them. 

\begin{figure}
 \centering
 \includegraphics[width=\columnwidth]{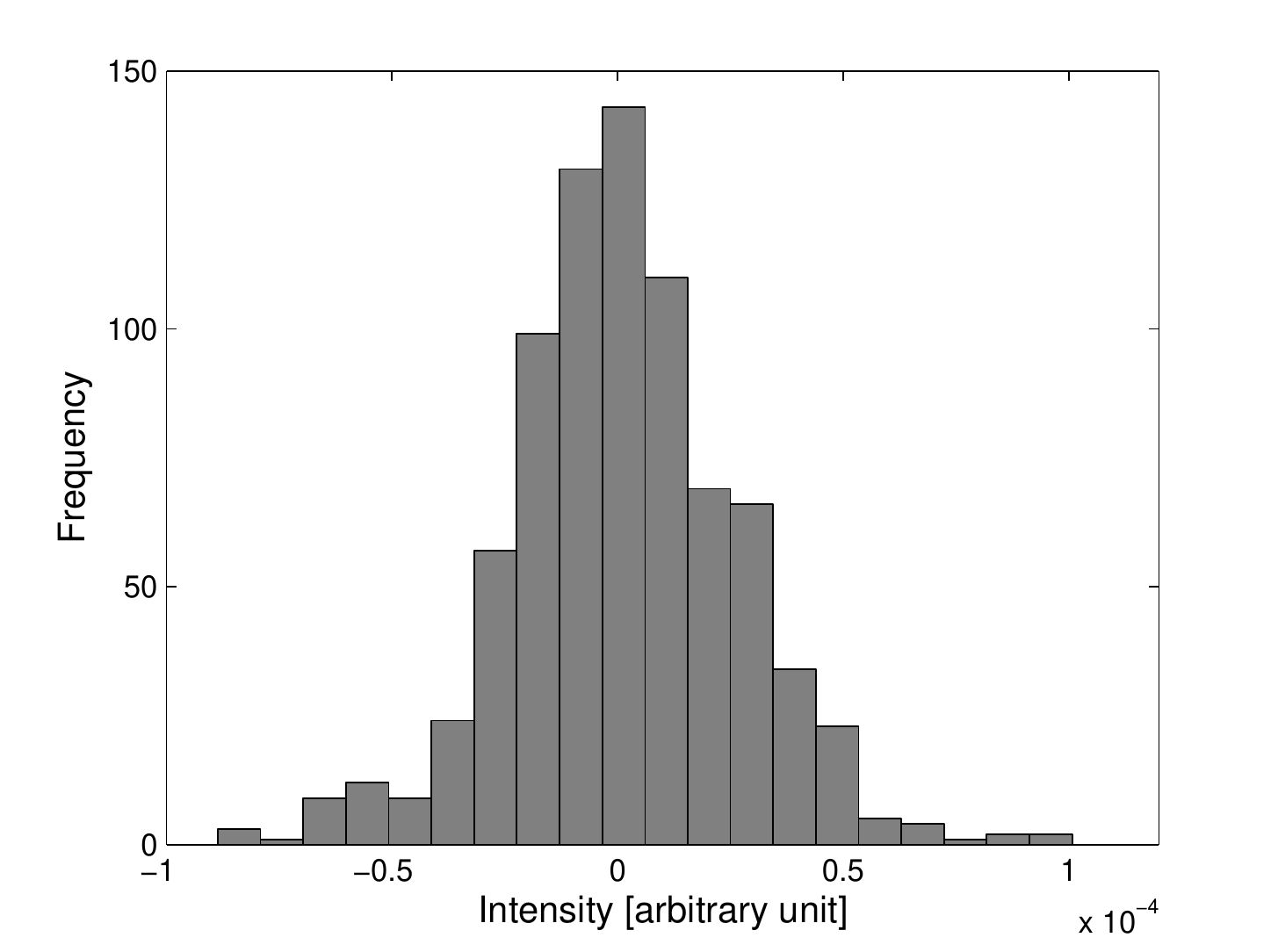}
 \caption{Intensity histogram computed from figure~\ref{fig:diffima} over a 4 pixel wide annulus of radius 0.8\arcsec{} containing 804 pixels.}
 \label{fig:specklenoise}
\end{figure}

Residual speckle noise does not necessarily follow a Gaussian distribution. Hence, error levels derived from a Gaussian confidence levels may not be appropriate. Chebyshev's inequality for example provides a lower significance limit of only 96\% for a 5$\sigma$ detection for a worst case intensity distribution. Figure~\ref{fig:specklenoise} shows the histogram of all pixels intensities in an 0.8\arcsec{} radius and 4 pixels wide annulus around the center of figure~\ref{fig:diffima} without artificial planets containing 804 pixels. Although the distribution appears Gaussian a detailed analysis reveals extended wings. For example the probability that a pixel has intensity above the 3$\sigma$ level is 1.5\% which is significantly larger than the 0.3\% of a Gaussian. If we want to be 90\% confident that none of the 804 pixels is above our $n\sigma$-criterion, a single pixel must have a probability smaller than $1-0.9^{(1/804)} = 0.013$\% to exceed this $n\sigma$ intensity. A Gaussian would provide this confidence using a 4.65$\sigma$ criterion. Since the residual speckle intensities are not strictly Gaussian, we add a margin and use 5$\sigma$. 

\begin{figure}
 \centering
 \includegraphics[width=\columnwidth]{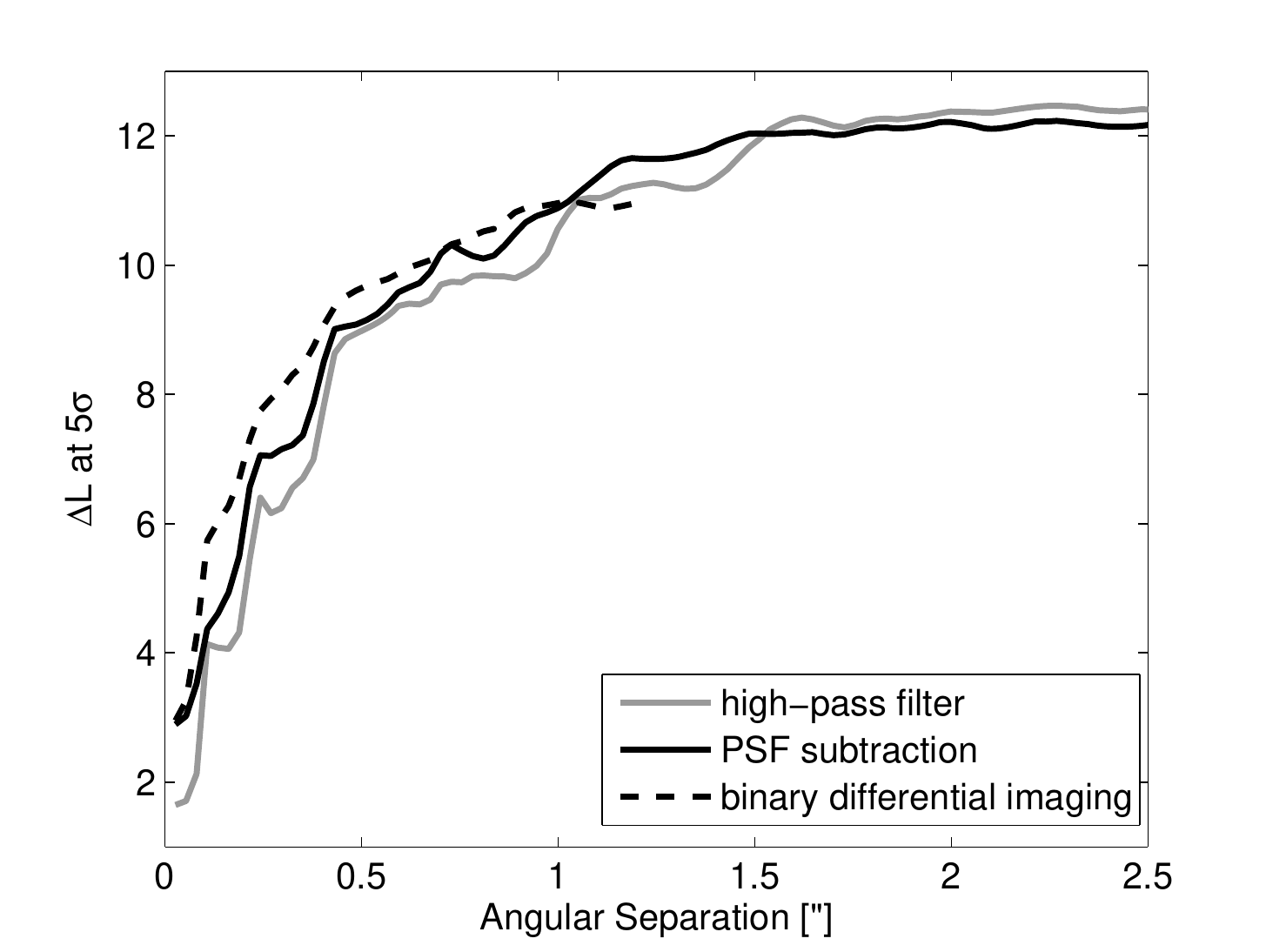}
 \caption{Contrast measured on the high-pass filtered image of HIP 102409 (gray solid line), the PSF subtracted image of HIP 102409 and HIP 107649 (black solid line) and the 2.8\arcsec{} binary HIP 102141 subtracting the components from each other (black dashed line). At separations larger than 1\arcsec{} to 1.5\arcsec, the achievable contrast is limited by sky background rather than stellar residual noise.}
 \label{fig:contrast}
\end{figure}

Figure~\ref{fig:contrast} plots the 5$\sigma$ contrast as a function of the angular distance for three different data types. The gray line represents the contrast obtained from the high-pass filtered image. About half to one magnitude can be gained by speckle calibration through PSF subtraction as shown by the black solid line representing data from the pair HIP 102409 and HIP 107649 shown in figures~\ref{fig:starfull} and \ref{fig:starhf}. The dashed line finally shows the contrast obtained by subtracting the two components of the 2.8\arcsec{} binary HIP 102141 from each other yielding an additional improvement of half to one magnitude. This line is representative of what could be achieved using differential techniques like L-band SDI. The flattening of the curves at larger angular distance is due to the sky background limited sensitivity of about L=16. 

The L--band imaging with NACO using either PSF subtraction or ADI typically achieves 5$\sigma$ contrast levels of about 9 magnitudes at 0.5\arcsec. This should be compared to H-band differential imaging delivering 10 magnitudes in the differential filter and 11.5 magnitudes assuming a T6 spectral type of the object \citep{biller07} and K-band PSF subtraction delivering 9 magnitudes contrast \citep{masciadri05}, both at 0.5\arcsec{} and 5$\sigma$ as well. Considering the H--L colors of around 2.5 to 4.5\,mag and even larger K--L for the objects we are looking for and our half a magnitude more pessimistic contrast estimation method, it becomes evident that the L-band imaging is currently the most sensitive method for the detection of close substellar companions that are cooler than about 1000\,K. This advantage does not hold for wide separation companions where L-band imaging cannot compete with shorter wavelength methods because of the reduced sky background limited sensitivity.

\section{Results}

\subsection{Discovery Space}
\begin{figure*}
 \centering
 \includegraphics[width=18cm]{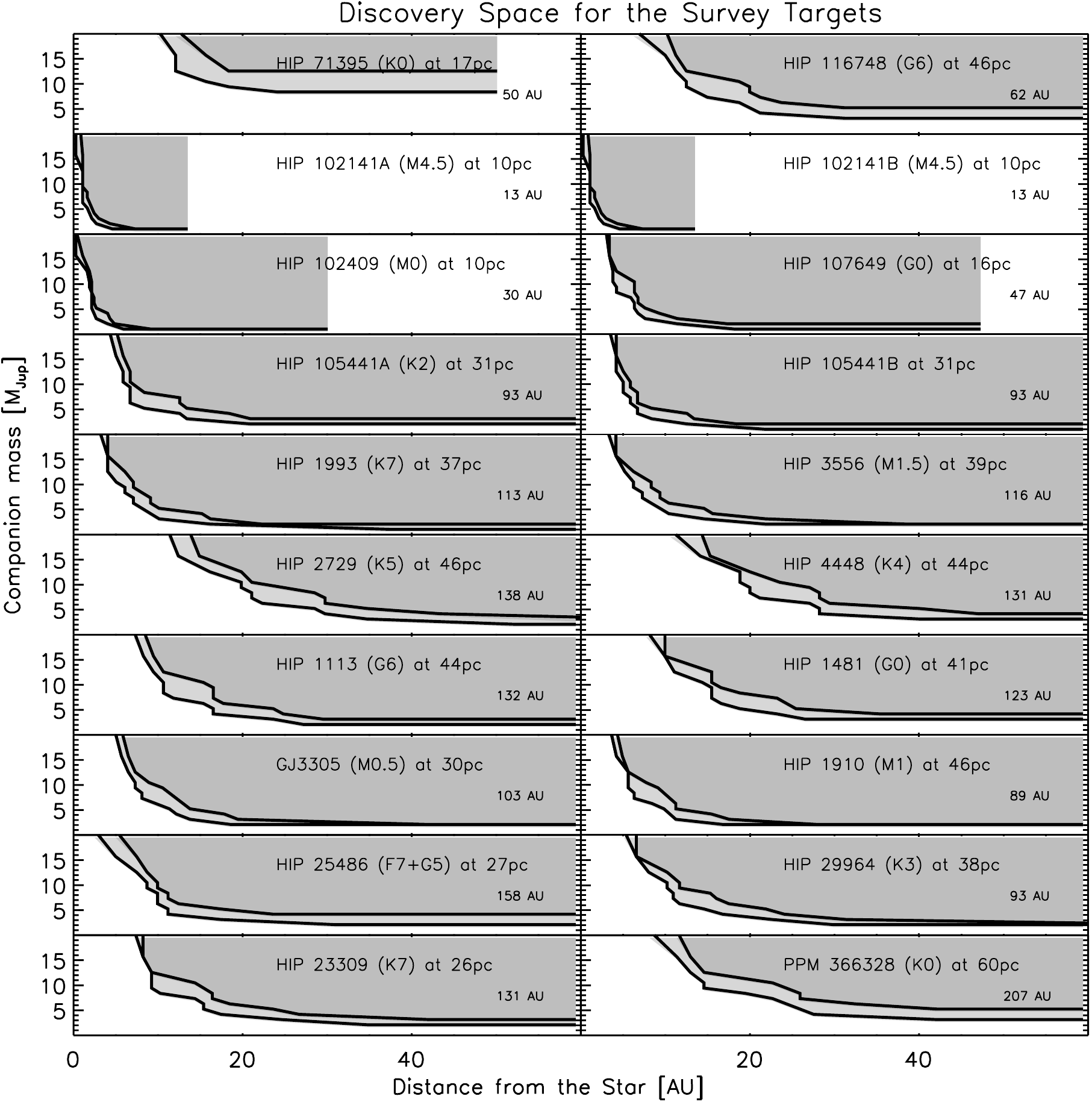}
 \caption{Discovery space for the observed targets for ages 10~Myr and 30~Myr, except for the older HIP 71395 (500 and 1000 Myr). The shaded area covers the planets that would have been detected by the survey. The field of view in AU is provided as well. The more sensitive lower curve is for the younger age.}
 \label{fig:ds}
\end{figure*}

Knowing the stellar ages, luminosities and distances (Table~\ref{tab:ts}), the calculated radial contrast for the individual objects can be converted into a discovery space using the planetary evolutionary models of \citet{baraffe03}. The discovery space describes the detectable planet mass as a function of separation from the star in AU and is shown for all target stars in figure~\ref{fig:ds}. Since no clear companion candidates were found, the discovery space provides upper limits on planet masses for a given separation from the host stars. Some of the observations are even sensitive to sub-Jupiter masses. Since these are not considered by the evolutionary models, we conservatively set 1\,M$_{\rm J}$ as the ultimate sensitivity limit.

\subsection{Uncertainty in the Evolutionary Models}
The dominant uncertainty in translating our sensitivity curves to planet masses is the uncertainty of the planet evolution models. Lacking direct detections of well--characterized young giant exoplanets, the simulated evolution at young ages is strongly influenced by the unknown initial conditions. The most widely used giant planet evolution models apply simple spherically contracting planetary atmospheres \citep[e.g.][]{baraffe03,burrows01,burrows04}. Efforts by different groups to include more realistic formation scenarios suggest a strong impact on the luminosity of young objects, but also suffer from the large number of unconstrained initial parameters \citep[e.g.][]{wuchterl03,marley07}. 
For instance, the recent simulations by \citet{marley07} incorporate a simplified core accretion formation scenario for giant planets. In this treatment a substantial fraction of the gravitational energy of the infalling gas is released in an energetic shock wave at the upper boundary of the planet's atmosphere, leading to the formation of giant planets up to several hundred times fainter than those predicted by spherically contracting models. In these models, the luminosity of the least massive planets deviates the least from the "hot start" models. \citet{marley07} suggest that a 1~M$_{\mathrm J}$ planet will "forget" its initial conditions by about 10 to 15 Myr, while for more massive planets this may take hundreds of Myr.

This comparison suggests that our observations are sensitive to 1--2~M$_{\mathrm J}$ planets around most target stars, independently of the models used. More massive planets are always brighter and would therefore have been confidently detected, but --- because their luminosity is highly model--dependent --- maybe on a lower significance level than suggested by the \citet{baraffe03} or \citet{burrows04} models.
 
\subsection{AU Mic}
Our target HIP 102409, better known as AU Mic, is of particular interest. This young, nearby M--dwarf has recently been identified to host a large ($r>210$~AU) edge--on debris disk. The surface density profile of the debris disks displays a dramatic break at 43~AU radius, suggesting a strong difference in the properties of the inner and outer disk. In addition, the lack of 10~$\mu$m excess emission from unresolved measurements hints on an inner hole with a few AU radii \citep{liu04}.  

The disk has been directly imaged using HST/ACS coronagraphy by \citet{krist05} as close to the star as 7.5~AU. The innermost disk does not display any significant structure or asymmetries that would indicate gravitational perturbation by a giant planet at those orbits. In accordance with this, our observations can exclude a 1~M$_{\mathrm J}$ planet at radii greater than 5~AU and less than 30~AU. 
 
\subsection{HIP 1910 and GJ 3305}
\begin{figure}
  \center{\includegraphics[width=2in,angle=0]{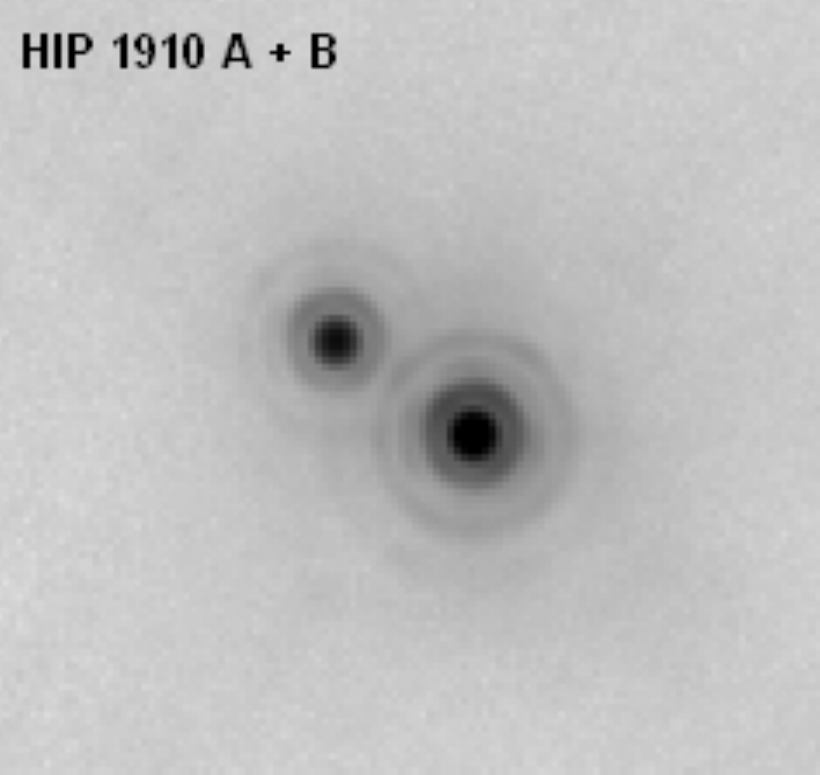}}
  \caption{NACO L-band images of the 0\farcs65 binary HIP 1910. North is up, East is left.}\label{fig:hip1910}
\end{figure}

\begin{figure}
  \center{\includegraphics[width=2in,angle=0]{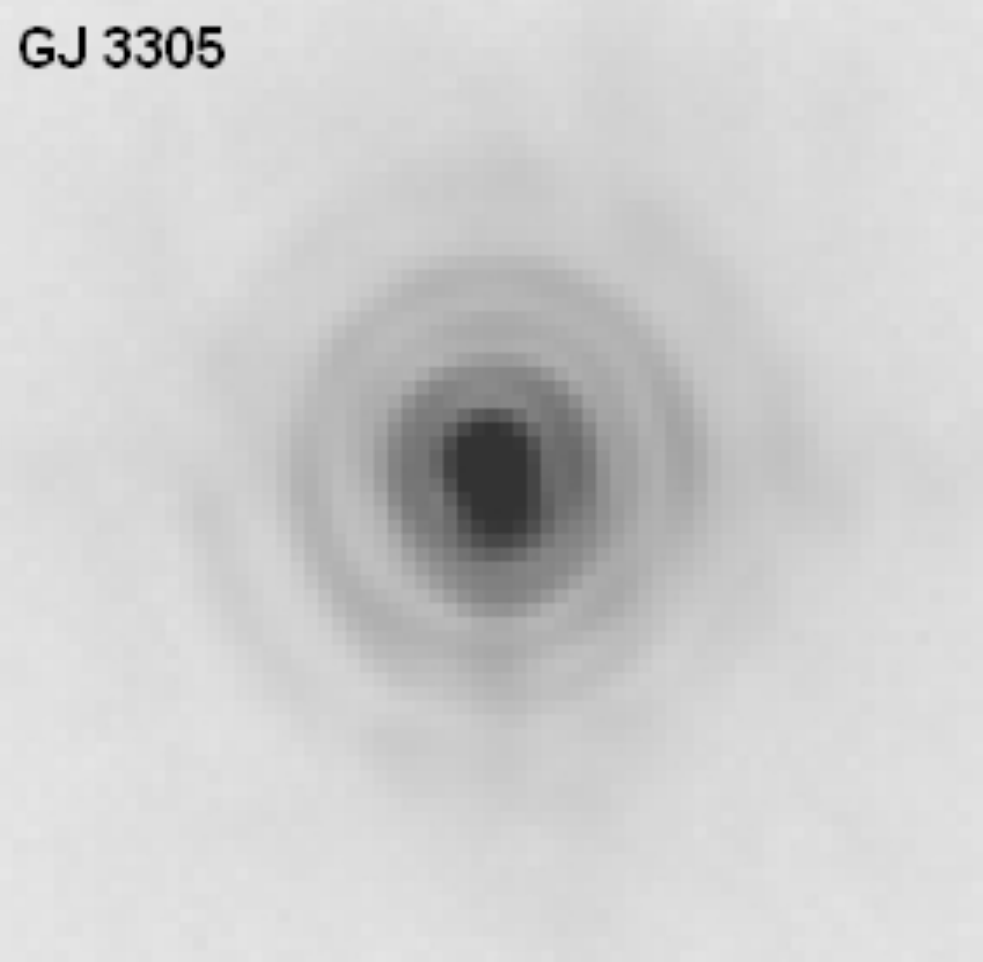}}
  \caption{NACO L-band images of the 0\farcs093 binary GJ 3305. North is up, East is left.}\label{fig:gj3305}
\end{figure}

Our data set contained two close binaries. The binarity of HIP~1910 (figure~\ref{fig:hip1910}) has been discovered by \citet{chauvin03}, while GJ 3305 (figure~\ref{fig:gj3305}) is a previously unknown binary if binarity can be confirmed by common proper motion. Astrometry and photometry of both binaries was performed by means of a one-dimensional least-squares fit of two model PSFs to the data along the binary's position angle. The results are presented in table\,\ref{tab:binprop}

\begin{table}
\caption{Astrometry and photometry of HIP~1910 (2 Dec. 2004) and GJ~3305 (15 Dec. 2004)}
\label{tab:binprop}      
\centering                          
\begin{tabular}{l l l l}        
\hline\hline                 
Source & L(mag) & Sep. (mas) & PA(\degr) \\    
\hline                        
   HIP 1910 A & $7.44\pm0.11$ & & \\      
   HIP 1910 B & $8.84\pm0.16$ & $647.6\pm0.5$ & 55.76$\pm$0.31 \\
   GJ 3305 A & $6.5\pm0.14$ & & \\
   GJ 3305 B & $7.38\pm0.24$ & $93.3\pm1.6$ & 189.5$\pm$0.4 \\
\hline                                   
\end{tabular}
\end{table}

Our NACO observations of HIP 1910 A/B combined with the results of the ADONIS observations by \citet{chauvin03} (2000 Nov 12: $698\pm29$\,mas separation, 49.6$\pm$2.5 degree PA, 2001 Oct 28: $699\pm30$\,mas, 50.2$\pm$2.5\,degree) yield clear evidence for the orbital motion of the binary. Over the 4 years between the first ADONIS observations and the NACO observations, the PA changed by 6.2+-2.5 degree, indicative of an orbital period of $\approx$230 years.

GJ~3305 and 51~Eri form a binary with a projected separation of 66\arcsec{} \citep{feigelson06}. The location of GJ~3305 in the HR-diagram and its activity (chromospheric and X-ray emission) let \citet{feigelson06} derive an age of about 13 Myr for the system. GJ~3305 was also observed by \citet{daemgen07} using ALTAIR at Gemini on Sep 29, 2005, and listed among the sources with no physical companion detection within the sensitivity limits. Our data shows GJ~3305~B at a position angle of 189.5 degrees and a projected separation of only about 93\,mas, i.e. 3\,AU at GJ~3305's distance of 30\,pc. Hence, orbital motion should be evident even within months, and the system mass could soon be determined from the orbit.

GJ~3305 has a proper motion of $\mu_{\rm RA} = 46.1 \pm 3.0$ mas/yr and $\mu_{\rm DEC} = -64.8 \pm 2.8$ mas/yr. If GJ~3305~B would be a background star, the separation between the two stars would have been different in the past. In fact, the ESO/ST-ECF Science Archive provides NACO K-band data of GJ~3305 from 18 January 2003. At this time GJ~3305~A and B were separated by 225\,mas at a position angle of 195 degrees, and GJ~3305~A was brighter in K-band by about 0.94\,mag. This result is inconsistent with GJ~3305~B being a background star, since the two stars would have appeared at 181\,mas separation and 239 degrees position angle at the time of our observations on 15 December 2004. The K and L-band photometry cannot discriminate between GJ~3305~B being an earlier type background star or an M-type companion within the measurement errors, but the flux measured on the AO visible wavefront sensor is consistent with a magnitude of V=10.6 for GJ~3305. Hence, GJ3305~B is as red as G~J3305~A providing additional strong evidence for the binary hypothesis. The orbit of GJ~3305 cannot be constrained much with the two data points, but it must be quite eccentric and currently close to periastron to provide the large change in position between January 2003 and December 2004. Such a change would not have occurred for a near circular orbit with a period of more than 17 years at the distance of 30\,pc.

\section{Statistical Analysis}

We use a Monte Carlo model to evaluate how our upper limits constrain the distributions of planet masses and orbital radii. 
In particular, we test which extrapolations of the exoplanet--populations detected by radial velocity methods is consistent or are in conflict with {\em only non--detections} in our sensitive survey. In the following we overview the model setup, the results of the simulations and discuss the statistical results.


\subsection{Model Setup}

The principle of our Monte Carlo simulation is to simulate a large number of planets consistent with different planet--mass and orbital radius distributions and compare them to our sensitivity limits. We will refer to a set of planets simulated around all observed stars as a {\em planet population}; planet--mass and orbital radius distributions to be tested are referred to as the {\em input parameter set}. In the course of the simulation for each input parameter set we generate 750 planet populations. Thus, for each input parameter set the ratio of planet populations without any planet detection over the total number of planet populations gives the probability that the input parameter set is consistent with only non--detections in our survey.

The major assumption of the simulation is that the probability distribution of giant planet mass and orbital radius are described by two 
independent  power--law functions ($P(r)\propto r^{\alpha}$ and $P(m)\propto m^{-1.5}$, between 0.3~AU and $r_{max}$, and 1~M$_{J}$ and 
50~M$_{J}$, respectively).  This simplification will certainly be refined in the next decade, but are likely to be a robust assumption to date.
 In our simulations we varied $\alpha$ and $r_{max}$. Lacking information on the eccentricity distribution of giant planets at larger orbital radii, we assumed circular orbits. Although adopting different eccentricity distributions has its merits, we argue that the detectability of a planet on eccentric orbit with a given semi--major axis would resemble statistically the detectability of a planet on circular orbit with the same orbital radius. The reason for this is that the key orbital parameter for the detectability is the planet--star separation: eccentric planets spend most of their time at large separations and only a very minor fraction close  to the star. Thus, in terms of our Monte Carlo
simulation the circular orbits provide a reasonable approximation that is not expected to influence the results.

In accordance with the age estimates for our targets, we assume 15 Myr for all but HIP 71395, which is estimated to be 500 Myr--old.

For each planet population we simulate planets around each of 20 observed stars (the very tight binaries HIP~1910 and GJ~3305 were not  treated separately). For each star we randomly determine whether it harbors planets, and if so, how many. We do this in such a way to keep the frequency of planets per star  consistent with the giant planet frequency determined from radial velocity surveys within the parameter space where these can be  considered complete (1--3 AU, $>1$M$_{J}$, $\sim$3\%, e.g. \citet{marcy05}). We chose the tested parameter space such to exclude the hot Jupiter population, as these obviously represent a group with a distinct orbital and mass distribution and cannot be tested through our observations. 

The probability that our observations are consistent with a certain parameter set is given by the fraction of populations with no planets detected. In other words, a parameter set in which 45\% of the simulated populations would have led to at least one planet detection with our sensitivity limits has a 45\% probability of being inconsistent with the measurements. Likewise, parameter sets that would result in planet detections in $>90\%$ of the simulations can be considered inconsistent with our measurements. The strength of this analysis lies in the identification of the parameter space that can be excluded based on our observations.

We note, that similar Monte Carlo simulations are being developed by several other groups. In particular, \citet{nielsen06} uses a similar  model to quantify the non--detections  from a spectral differential imaging survey \citep{biller07}, and Heinze et al. (in prep.) utilize this technique on an MMT planet survey. Apai et al. (in prep.) use a slightly modified version of the simulations presented here to interpret the outcome of their MMT/Clio L--band survey.

\subsection{The Modeling Process}

We define a regularly--spaced grid in the two--dimensional parameter space in $\alpha$ and $r_{max}$. For each grid point  we simulate 750 planet populations. For each of the 20 stars within these populations, the following
steps are taken:

(1) We calculate the {\em average} number of giant planets for the given $\alpha$ and $r_{max}$  values in such a way that the planet frequency in the RV--surveyed orbits is consistent with the observed frequency, as outlined in the model setup.

(2) Using a random number generator and the average number of giant planets we determine the {\em actual} number of giant planets around this specific star.

(3) Using the $P(m)\propto m^{-1.5}$ and $P(r)\propto r^{\alpha}$ power--laws we determine randomly the mass and orbital radius of each giant planet around the star.  We poll a random orbital phase and an inclination. If there are multiple planets in a system, we use the same inclination for all.

(4)  Based on the distance of the star and the planet's orbital radius, inclination, and phase angle we determine the apparent planet--star separation. By comparing the system age and the planet mass to the \citet{baraffe03} isochrones and by applying the distance modulus, we calculate the apparent magnitude of each planet. 

(5) We compare the apparent separation and apparent magnitude of the planet to the sensitivity curve measured for the given star and determine whether or not the planet would have been detected by our measurements.

The key result for each grid point is the ratio of the number of populations with at least one planet detected over the total number of populations, i.e. the frequency of simulated "surveys" that are inconsistent with our real survey.

\subsection{Statistical Results and Discussion}
 
\begin{figure*}
 \centering
 \includegraphics[width=12cm]{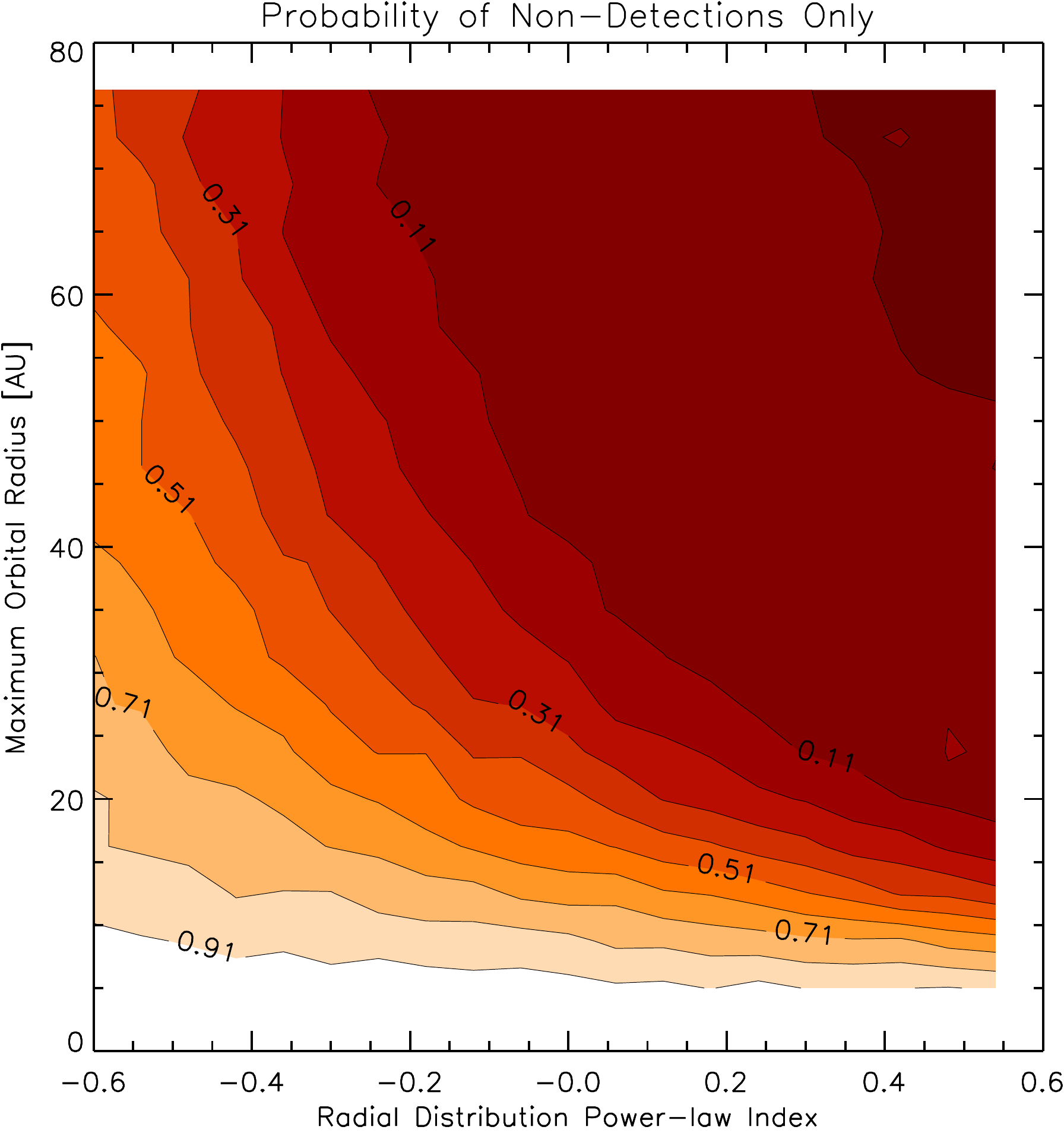}
 \caption{Map of probability that the planet population simulated for a given $\alpha$ and $r_{max}$ value is consistent with the non--detections in our survey.}
 \label{fig:grid}
\end{figure*}

Fig.~\ref{fig:grid} shows the result of the Monte Carlo simulation. In this two--dimensional probability map the value of the map for any given $\alpha$ and $r_{max}$ values gives the probability that these planet mass and orbital radius distributions are consistent with {\em only} non--detections in our survey. The dark--shaded region with probabilities below 0.1 ($\alpha \ge 0.2$ and $r_{max} \ge 30$~AU or $\alpha \ge 0$ and $r_{max} \ge 40$~AU) can be excluded at a confidence level of 90\%; we believe that planet distributions with these parameters would have been detected by our observations.  It is interesting to note that this simple analysis shows that the more likely distributions of the giant planet population have small outer radii ($r_{max} < 15$~AU) and a decreasing radial distribution ($\alpha < 0$).

Such a distribution is somewhat expected --- although by no means proven --- on the basis of the architecture of the Solar System. Most current planet formation models focus on explaining the inner planetary systems constrained through radial velocity, gravitational lensing and transiting planet searches. Very few, if any, predictions are made on the outer planetary systems. Most models artificially truncate the protoplanetary disks at a somewhat arbitrary radius. Our survey demonstrates that current high-contrast imaging can place meaningful constraints on the outer edge of the planet population.  

\section{Conclusions}
L-band observations with NACO provide high-contrast, high Strehl ratio (up to 70\%) data under median seeing conditions. This survey technique is complementary to Spectral Differential Imaging (SDI), as i) cool substellar companions tend to be brighter while the host star gets fainter at longer wavelength, and ii) unlike SDI, the L-band observations do not rely on the presence of strong spectral features like Methane in the atmospheres of substellar or planetary mass companion. In general, our adaptive optics L-band survey was sensitive to planets with masses down to 1 to 2 M$_{\rm J}$ at separations larger than 5 to 30 AU around the host stars. The presented technique is currently the most sensitive method for the contrast limited detection of substellar companions that are cooler than about 1000\,K.

We can exclude the presence of a planet with a mass $\ge$1 M$_{\rm J}$ at distances $>$5 AU and $<$30 AU around AU Mic. The abrupt change in disk morphology around AU Mic at a radius of $\approx$7.5\,AU\ does not seem to be related to the presence of the giant planet close to the inner edge of the disk.

GJ\,3305 was resolved into a very close ($\approx$93\,mas) binary. The proximity of the two components and the quick change in orbital position should allow one to measure the orbit and determine the system mass of this young object within a few years.

Based on our L-band survey, the frequency of giant planets with masses above 2-3 M$_{\rm J}$ at separations larger than 30 AU around nearby, G, K and M stars is less than about 5\%. If we combine this with the results by \citet{masciadri05}, who surveyed 28 nearby stars that partly (6 stars) overlapped with our sample, and \citet{brandner00}, who surveyed 24 M-stars in nearby star forming regions, the frequency of brown dwarf companions to low-mass stars at separations $\ge$30\,AU\ is less than 1\%, and the frequency of giant planets with masses $\ge$5\,M$_{\rm J}$ is less than 2\%. Our Monte-Carlo simulations indicate that our results are not consistent with a power-law of planet frequency exceeding $\alpha\ge0-0.2$ and a maximum planet separation of $r_{max}\ge30-40$~AU at a confidence level of 90\%.

\begin{acknowledgements}
We thank the ESO VLT astronomers for carrying out our observations in service mode and Ben Zuckerman for helpful comments on the manuscript. This material is partly based upon work supported by the National Aeronautics and Space Administration through the NASA Astrobiology Institute under Cooperative Agreement No. CAN-02-OSS-02 issued through the Office of Space Science. WB acknowledges support by a Julian Schwinger fellowship through UCLA.
\end{acknowledgements}

\bibliographystyle{aa}

\bibliography{references}

%
%
%
%
%
%
%
%
%

\end{document}